\documentclass[aps,prl,reprint,nofootinbib,superscriptaddress]{revtex4-2}
\usepackage[T1]{fontenc}
\usepackage{lmodern}
\usepackage{amsmath,amssymb,bm}
\usepackage{graphicx}
\usepackage{booktabs}
\usepackage{microtype}
\usepackage[
  colorlinks=true,
  allcolors=blue,
  pdftitle={High-Power AM-CW Lunar Laser Ranging as a $\mu$Hz SGWB Detector},
  pdfauthor={Slava G. Turyshev}
]{hyperref}
\newcommand{\Ogw}{\Omega_{\rm gw}}
\newcommand{\hc}{h_{\rm c}}
\newcommand{\uHz}{\mu{\rm Hz}}
\newcommand{\mum}{\,\mu{\rm m}}
\newcommand{\mums}{\,\mu{\rm m}\,{\rm s}^{-1}}
\newcommand{\kmsmpc}{\mathrm{km\,s^{-1}\,Mpc^{-1}}}
\newcommand{\fii}{f_2}
\newcommand{\Dcov}{\mathcal{D}_{\rm cov}}
\newcommand{\Dnorm}{\mathcal{D}_{\rm norm}}
\newcommand{\Dcamp}{\mathcal{D}_{\rm camp}}
\newcommand{\Dtot}{\mathcal{D}_{\rm tot}}
\newcommand{\Did}{\mathcal{D}_{\rm id}}
\newcommand{\Dmax}{\mathcal{D}_{\rm tot}^{\max}}
\newcommand{\nuEff}{\nu_{\rm eff}}
\newcommand{\LEM}{L_{\rm EM}}

\begin{document}
\setlength{\bibsep}{0pt}

\title{High-Power AM-CW Lunar Laser Ranging as a $\mu$Hz SGWB Detector}

\author{Slava G. Turyshev}
\affiliation{Jet Propulsion Laboratory, California Institute of Technology, Pasadena, California 91109, USA}

\date{September 6, 2026}

\begin{abstract}
High-power amplitude-modulated continuous-wave lunar laser ranging (AM-CW LLR) is a stochastic-gravitational-wave-background (SGWB) experiment that measures GW-driven drift and diffusion of the Earth--Moon orbit.  Its observable is the coherent phase of a GHz modulation envelope returned by an individually resolved single hollow corner-cube retroreflector (HCCR), converted to absolute range and analyzed through its time-dependent mean and covariance.  The response occurs at lunar orbital harmonics; the dominant low-eccentricity channel is centered at $f_2=0.847245\,\mu{\rm Hz}$, and a five-year broadband forecast applies to spectra smooth over the resolution scale $T_{\rm obs}^{-1}=6.34\,$nHz about that resonance.  A scalar 1-kW, 1064-nm link yields $5.893\times10^3$ detected photons s$^{-1}$ and an $88.8$--$124.3\,\mu$m photon contribution in 100 s, while the forecasts adopt projected total range precisions of $80$ and $50\,\mu$m.  In the covariance-dominated weak-signal regime, $\Omega_{\rm gw}^{95}\propto\sigma_R^2\nuEff^{-1}T_{\rm obs}^{-4}$.  Using the published Earth--Moon response normalization, the five-year 95\% thresholds are $1.86\times10^{-7}\Dtot$ for $80\,\mu$m and $500\,{\rm yr}^{-1}$ and $2.42\times10^{-8}\Dtot$ for $50\,\mu$m and $1500\,{\rm yr}^{-1}$, where $\Dtot$ is the full-analysis threshold divided by the unit-factor response-anchored forecast.  If end-to-end calibration, sampling, and marginalization give $\Dtot\le1.15$, the higher-performance campaign reaches nominal source-statistic equality for $\Omega_{\rm gw}=7.08\times10^{-8}$ in five years.  These measurable performance and information-retention requirements define high-power AM-CW LLR as a standalone quantitative SGWB experiment at the $0.85\,\mu{\rm Hz}$ Earth--Moon resonance.
\end{abstract}

\maketitle

\emph{Motivation.}--A stochastic gravitational-wave background in the $10^{-7}$--$10^{-4}\,{\rm Hz}$ band can contain peaks, turnovers, or local structure from first-order phase transitions, cosmic strings, blue-tilted primordial tensor spectra, environmentally coupled compact binaries, and massive-black-hole environments \cite{Sesana2021,BlasJenkinsPRL2022,Regimbau2011,CapriniFigueroa2018,Auclair2020}.  The Earth--Moon binary is a resonant dynamical detector in this band: GW tidal forcing produces drift and diffusion of the lunar orbital elements, and lunar laser ranging (LLR) converts that response into a measured time series of absolute ranges.  Here we formulate high-power amplitude-modulated continuous-wave \mbox{(AM-CW)} LLR as an SGWB experiment and derive its measurement principle, orbital-harmonic frequency response, five-year sensitivity, and commissioning requirements.  We use the published Earth--Moon response ordinate to normalize the orbital dynamics while deriving the AM-CW observable, noise model, sampled response, and likelihood independently.  The analysis thereby specifies an experiment-level constraint and detection criterion rather than inferring performance from range precision alone.

Three levels of result are kept separate.  Equations~(\ref{eq:range_mean})-- (\ref{eq:range_kernel}) and Eqs.~(\ref{eq:phase})-- (\ref{eq:shot}) establish the SGWB-response and measurement frameworks, respectively.  The scalar link rate, the $80$ and $50\,\mu$m error budgets, and the accepted-window cadences are projected engineering cases.  An experiment-level sensitivity additionally requires the measured HCCR transfer $\eta_H(\lambda,\alpha_{\rm va},\iota,T_H)$, realized timestamps, correlated measurement covariance, and sampled drift-and-diffusion kernel in the global lunar fit.  The numerical curves below are requirements on those data products rather than sensitivity inferred from range precision alone.

A Gaussian SGWB does not appear as one deterministic sinusoid in a range residual.  Stochastic forcing can induce both a noise-averaged drift and a covariance of the fitted Earth--Moon orbit; the covariance is its distinctive zero-mean fluctuation channel.  The source quantity is
\begin{equation}
 \Ogw(f)=\frac{1}{\rho_c}\frac{d\rho_{\rm gw}}{d\ln f},\qquad
 \rho_c=\frac{3H_0^2c^2}{8\pi G} .
 \label{eq:OmegaDef}
\end{equation}
Here $\rho_c$ is the critical energy density and $H_0$ is the Hubble constant. In a local transverse-traceless (TT) frame the tidal acceleration of a binary separation $\bm r$ is
\begin{equation}
 a^i_{\rm gw}=\frac{1}{2}\ddot h^{\rm TT}_{ij}r^j .
 \label{eq:tidal}
\end{equation}
For a Gaussian SGWB, the osculating elements $X^i=\{P,e,I,\Omega,\omega,\epsilon\}$ acquire an SGWB-induced Fokker--Planck drift increment and diffusion coefficients
\begin{align}
 \delta D^{(1)}_i(X)&=\sum_{n\ge1} A_{n,i}(X)\,\Ogw(n/P),\nonumber\\
 D^{(2)}_{ij}(X)&=\sum_{n\ge1} B_{n,ij}(X)\,\Ogw(n/P),
 \label{eq:diffusion}
\end{align}
so the response is at orbital harmonics \cite{BlasJenkinsPRL2022,BlasJenkinsPRD2022,Foster2025,FosterMethods2025}. Here $P,e,I,\Omega,\omega,\epsilon$ denote orbital period, eccentricity, inclination, longitude of ascending node, argument of periapsis, and orbital phase, while $A_{n,i}$ and $B_{n,ij}$ are the sky- and polarization-averaged drift and diffusion response coefficients.  In Fokker--Planck notation,
\begin{equation*}
 D_i^{(1)}\equiv\lim_{\Delta t\to0}\frac{\langle\Delta X^i\rangle}{\Delta t},\qquad
 D_{ij}^{(2)}\equiv\frac12\lim_{\Delta t\to0}
 \frac{\langle\Delta X^i\Delta X^j\rangle_c}{\Delta t}.
\end{equation*}
The ordinary deterministic drift $V_i$ remains in the nominal ephemeris, so $D_i^{(1)}=V_i+\delta D_i^{(1)}$, while $D_{ij}^{(2)}$ is the orbital-element diffusion tensor and the incremental second moment grows as $2D_{ij}^{(2)}dt$.  These dimensional roman-$D$ coefficients generate the physical mean and covariance response; they are unrelated to the dimensionless calligraphic-$\mathcal D$ sensitivity factors introduced below.

The connection to the measured light time is obtained by linearizing the global range model about its nominal solution.  For observation $k$,
\begin{equation}
 \delta R_k^{\rm GW}=J_{ki}\,\delta X^i(t_k),\qquad
 J_{ki}=\left.\frac{\partial R(t_k,\bm X)}{\partial X^i}
 \right|_{\bm X_0} ,
 \label{eq:range_response}
\end{equation}
and hence, for response functions ${\cal M}_k$ and ${\cal K}_{kl}$,
\begin{align}
 \left\langle\delta R_k^{\rm GW}\right\rangle
 &=\int d\ln f\,\Ogw(f)\,{\cal M}_k(f),\label{eq:range_mean}\\
 {\rm Cov}(\delta R_k^{\rm GW},\delta R_l^{\rm GW})
 &=\int d\ln f\,\Ogw(f)\,{\cal K}_{kl}(f).
 \label{eq:range_kernel}
\end{align}
The functions ${\cal M}_k$ and ${\cal K}_{kl}$ contain the sky- and polarization-averaged drift and tidal diffusion, propagation of the perturbed osculating elements, two-way light-time partials $J_{ki}$, and the sampling window.  This completes the physical chain $h_{ij}\rightarrow a_{\rm gw}\rightarrow\delta X^i\rightarrow\delta R \rightarrow\{\mu_R^{\rm GW},C_R^{\rm GW}\}$; it is sharply enhanced near orbital harmonics but is not a deterministic range sinusoid.

For the Moon, the sidereal orbital period and second-harmonic frequency are
\begin{equation}
 P_{\rm M}=27.321661\,{\rm d},\qquad
 \fii=2/P_{\rm M}=0.847245\,\uHz,
 \label{eq:f2}
\end{equation}
and the $n=2$ harmonic dominates the low-eccentricity response.  Its gravitational wavelength and long-wavelength ratio are
\begin{equation}
 \lambda_{\rm gw}=\frac{c}{\fii}=3.54\times10^{14}\,{\rm m},\qquad
 \frac{\LEM}{\lambda_{\rm gw}}=1.09\times10^{-6}.
 \label{eq:longwavelength}
\end{equation}
The Earth--Moon system is therefore safely in the tidal regime of Eq.~(\ref{eq:tidal}).  The direct light-propagation perturbation is of order $h\LEM$ and does not accumulate resonantly; the multi-year sensitivity below comes instead from the GW-driven orbital response.  The equivalent characteristic strain is \cite{Maggiore2000}
\begin{equation}
 \hc(\fii)=\left[\frac{3H_0^2}{2\pi^2}\frac{\Ogw(\fii)}{\fii^2}\right]^{1/2}
 =3.19\times10^{-17}\left(\frac{\Ogw}{10^{-9}}\right)^{1/2},
 \label{eq:hc}
\end{equation}
using $H_0=67.66\,\kmsmpc$ \cite{Planck2020}.  With the mean Earth--Moon separation $\LEM=3.844\times10^8\,$m, this gives the dimensional scale
\begin{equation}
 \LEM\hc(\fii)=12.3\,{\rm nm}
 \left[\frac{\Ogw(\fii)}{10^{-9}}\right]^{1/2}.
 \label{eq:lengthscale}
\end{equation}
A unit-projection transverse--traceless arm response would have the characteristic scale $\hc\LEM/2=6.1\,$nm per logarithmic frequency interval. These are spectral normalization scales, not an instantaneous range realization or a secular displacement to be quoted in distance per year.  A finite-band root-mean-square response also contains the bandwidth and antenna response.  Although the strain has zero ensemble mean, noise-induced orbital drift need not vanish; the measured signature is the resonantly accumulated drift and covariance in Eqs.~(\ref{eq:range_mean}) and (\ref{eq:range_kernel}).

\emph{AM-CW phase-ranging observable.}--The AM-CW LLR detector measures the radio-frequency (RF) envelope phase on a bright continuous optical carrier and converts it to the one-way geometric range $R=c\tau/2$ inferred from the two-way light time $\tau$ \cite{TuryshevCW2025,TuryshevAMCW2025}.  This is a coherent phase observable rather than an average of pulsed photon arrival times.  The architecture uses passive, individually resolved single corner-cube retroreflectors (CCRs) at distributed lunar sites.  For a legacy array with one-way cube offsets $\delta R_j$ and photon weights $w_j(t)$, the target transfer is
\begin{equation}
 H_{\rm arr}(f_m,t)=
 \frac{\sum_jw_j(t)e^{-i4\pi f_m\delta R_j(t)/c}}
      {\sum_jw_j(t)} .
 \label{eq:arraytransfer}
\end{equation}
Libration changes both $|H_{\rm arr}|$ and its phase.  A modern single hollow corner-cube retroreflector (HCCR) would remove this array-depth transfer and bulk-glass path, but not velocity aberration, dihedral error, polarization/coating phase, support motion, or a thermoelastic phase center.  Its vector far field and phase center must therefore be qualified; an HCCR is not assigned a photon gain from aperture alone \cite{TuryshevHCCR2013,PrestonMerkowitz2013,WilliamsCCR2023, TuryshevHCCR2026}.

For modulation frequency $f_m$,
\begin{equation}
 P_{\rm tx}(t)=P_0[1+a_m\cos(2\pi f_mt)] ,
 \label{eq:transmit}
\end{equation}
where the subscript ``tx'' denotes transmit, $P_0$ is the mean transmitted optical power, and $a_m$ is the transmitted modulation depth.  A scalar screening link is
\begin{equation}
 \dot N_s=\frac{P_0}{E_\gamma}\,\eta_{\rm eff}\eta_H
 \frac{A_H}{\pi(\LEM\theta_{\rm tx})^2}
 \frac{A_{\rm rx}}{\pi(\LEM\theta_{\rm ret})^2}.
 \label{eq:link}
\end{equation}
Here $E_\gamma=h_{\rm P}c/\lambda$ is the photon energy and $h_{\rm P}$ is Planck's constant.  The subscripts ``rx'' and ``ret'' denote receiver and return; $D_{\rm tx}$, $D_{\rm rx}$, and $D_H$ are the transmitter, receiver, and single-HCCR clear-aperture diameters, and $A_H=\pi D_H^2/4$ and $A_{\rm rx}=\pi D_{\rm rx}^2/4$ are the corresponding HCCR and receiver areas.  These roman-$D$ aperture symbols enter the optical link and are unrelated to the calligraphic-$\mathcal D$ sensitivity factors.  For the scalar reference screen, the effective uplink and HCCR first-null half-angles are
\begin{align*}
 \theta_{\rm tx}&=\left[\left(\frac{1.22\lambda}{D_{\rm tx}}\right)^2
 +\left(\frac{\lambda}{r_0}\right)^2\right]^{1/2}
 =5.476\,\mu{\rm rad},\\
 \theta_{\rm ret}&=\frac{1.22\lambda}{D_H}=12.981\,\mu{\rm rad},
\end{align*}
for $\lambda=1064\,$nm, $D_{\rm tx}=D_{\rm rx}=1\,$m, $r_0=0.20\,$m, and $D_H=0.10\,$m \cite{TuryshevCW2025}, where $r_0$ is the Fried atmospheric coherence parameter at 1064 nm.  The coefficient 1.22 locates the first Airy null of a circular aperture, whereas $\lambda/r_0$ is the conventional Fried coherence-angle estimate adopted in this scalar screen; the different prefactors therefore represent different angular conventions.  Their quadrature combination is an engineering envelope, not a matched point-spread-function convolution.  Thus $\theta_{\rm tx}$ is a diffraction-plus-seeing design angle rather than the diffraction limit of the 1-m transmitter.  The adopted scalar efficiency is
\begin{align*}
 \eta_{\rm eff}&=\eta_{\rm tel}^2\eta_{\rm atm}^2
 \eta_{\rm refl,0}\eta_{\rm ab,0}\eta_{\rm det}\\
 &=(0.80)^2(0.85)^2(0.70)(0.75)(0.80)\\
 &=0.194208\simeq0.194,
\end{align*}
where $\eta_{\rm tel}$ and $\eta_{\rm atm}$ are the single-pass telescope and atmospheric transmissions, $\eta_{\rm refl,0}$ is the nominal reflector efficiency, $\eta_{\rm ab,0}$ is the nominal velocity-aberration coupling, and $\eta_{\rm det}$ is the detector efficiency.  The squared telescope and atmospheric factors represent the outgoing and return paths.  The numerical factors follow the reference design assumptions of Refs.~\cite{TuryshevCW2025,TuryshevAMCW2025}.  These component efficiencies are projected values, not measured lunar HCCR throughput.  The factor $\eta_H(\lambda,\alpha_{\rm va},\iota,T_H)$ is a dimensionless HCCR throughput ratio to be measured and qualified.  It is normalized to unity at the same nominal geometry and thermal state $(\alpha_{{\rm va},0},\iota_0,T_{H,0})$ represented by the reference factors above; its measured replacement must not reintroduce the reflector or nominal aberration factors already contained in $\eta_{\rm eff}$.  Here $\alpha_{\rm va}$ is the velocity-aberration angle, $\iota$ is the incidence angle, and $T_H$ is the HCCR thermal state.  With $P_0=1\,$kW and $\eta_H=1$, Eq.~(\ref{eq:link}) gives the scalar reference normalization $\dot N_s=5.893\times10^3\,$s$^{-1}$.  The geometric and end-to-end fractions are $-172.5$ and $-179.6\,$dB.  For comparison, an unoptimized circular scalar far field has
\begin{equation*}
 \eta_{\rm FF}(\alpha_{\rm va})=
 \left[\frac{2J_1(x)}{x}\right]^2,
 \qquad x=\frac{\pi D_H\alpha_{\rm va}}{\lambda},
\end{equation*}
where $J_1$ is the first-order Bessel function.  The scalar response falls from 0.698 to 0.200 over $\alpha_{\rm va}=4$--$8\,\mu{\rm rad}$.  Relative to the nominal 0.75 aberration factor already included in $\eta_{\rm eff}$, this corresponds to $\eta_H=0.93$--0.27 and $\dot N_s=5.5\times10^3$-- $1.6\times10^3\,$s$^{-1}$ when the other HCCR variables remain at their reference values.  A measured vector far field and the realized geometry distribution replace this scalar illustration.  A 2-m receiver and $\eta_{\rm eff}=0.35$ give $4.25\times10^4\,$s$^{-1}$ before the measured factor $\eta_H$ is applied.  All of these rates are scalar model projections, not demonstrated lunar HCCR returns.

The observed return phase from target $a$ is
\begin{equation}
 \phi_a^{\rm obs}(t)=\frac{4\pi f_m}{c}R_a(t)
 +\phi_{\rm inst}(t)+\delta\phi_{H,a}(t)+\eta_{\phi,a}(t),
 \label{eq:phase}
\end{equation}
where $\delta\phi_{H,a}$ contains target- and geometry-dependent terms and $\eta_{\phi,a}$ is photon noise.  Neither $\phi_{\rm inst}$ nor $\delta\phi_{H,a}$ is assumed to vary rapidly or disappear on averaging. With $\phi_a^{\rm cal}=\phi_a^{\rm obs}-\widehat\phi_{\rm inst} -\widehat{\delta\phi}_{H,a}$, and holding $f_m$ fixed within a precision window, the calibrated mean and slope give
\begin{equation}
 \widehat R_a=\frac{c}{4\pi f_m}
 (\overline{\phi}_a^{\rm cal}-\phi_0+2\pi N),\qquad
 \widehat v_{r,a}=\frac{c}{4\pi f_m}
 \frac{d\phi_a^{\rm cal}}{dt},
 \label{eq:estimators}
\end{equation}
where $\phi_0$ is the calibrated phase origin and the integer $N$ is removed with multi-tone synthetic wavelengths \cite{TuryshevAMCW2025}.  Residual means and slopes of the two subtracted phase terms bias $R$ and $v_r$; their stochastic parts remain in the measurement covariance.  For a sufficiently slow deliberate slew, the leading-order rate correction is
\begin{equation}
 \widehat v_{r,a}=\frac{c}{4\pi f_m}\frac{d\phi_a^{\rm cal}}{dt}
 -\widehat R_a\frac{\dot f_m}{f_m}.
 \label{eq:slew_corrected_rate}
\end{equation}
The exact demodulator uses the recorded phase-history difference $\Phi_m(t)-\Phi_m(t-\tau)$; for a linear chirp this also contains the familiar $-\pi\dot f_m\tau^2$ term.  Thus the approximation above is not used in place of the phase history.  If the slew were left uncompensated, a $1\mum\,{\rm s}^{-1}$ rate allocation at $\LEM$ would require $|\dot f_m|<2.6\times10^{-6}\,{\rm Hz\,s^{-1}}$ for $f_m=1\,$GHz.  At that modulation frequency, the single-tone ambiguity interval is $c/(2f_m)=0.150\,$m.  A second tone separated by $\Delta f$ supplies the synthetic interval $c/(2\Delta f)$ (for example, $1.50\,$km at $\Delta f=0.1\,$MHz), with the modeled light time selecting the remaining coarse branch.  The corresponding phase-to-range scale is
\begin{equation}
 \delta\phi=4.19\times10^{-3}
 \left(\frac{\delta R}{100\mum}\right)
 \left(\frac{f_m}{1\,{\rm GHz}}\right){\rm rad}.
 \label{eq:phasescale}
\end{equation}

Photon-counting AM-CW does not reconstruct every RF cycle.  Time-tagged arrivals are folded against a maser-referenced, Doppler-derotated local oscillator (LO),
\begin{equation}
 Z_m(T_{\rm int})=\sum_{n=1}^{N_{\rm det}}e^{-i\Phi_{\rm LO}(t_n)} .
 \label{eq:phasor}
\end{equation}
Here $N_{\rm det}$ is the number of detected photon events in an integration window of duration $T_{\rm int}$.  We call the calibrated range estimate and covariance formed from one accepted $T_{\rm int}$ window an AM-CW normal-point equivalent; it combines many time-tagged photon events and is not an individual photon return.  For signal and background rates $\dot N_s$ and $\dot N_b$, a conservative phasor signal-to-noise ratio (SNR) design expression is
\begin{align}
 {\rm SNR}_{\rm AM}&\simeq\frac{a_{m,{\rm eff}}}{2}
 \left[\frac{\dot N_sT_{\rm int}}
 {1+\dot N_b/\dot N_s}\right]^{1/2},\nonumber\\
 \sigma_{R,{\rm shot}}&\simeq
 \frac{c}{2\pi f_ma_{m,{\rm eff}}}
 \left[\frac{1+\dot N_b/\dot N_s}
 {\dot N_sT_{\rm int}}\right]^{1/2}.
 \label{eq:shot}
\end{align}
The range expression is a factor $\sqrt{2}$ above the ideal sinusoidal Poisson phase Cram\'{e}r--Rao bound and is retained as a design allocation, not asserted as a fundamental photon limit. Here $a_{m,{\rm eff}}$ includes the transmitter, receiver, and event-timing transfer; Gaussian timing jitter $\sigma_t$ multiplies it by $\exp[-(2\pi f_m\sigma_t)^2/2]$ (0.82 at 1 GHz for $100\,$ps). A $1\,$km s$^{-1}$ line-of-sight speed shifts the two-way 1-GHz envelope by $6.67\,$kHz, hence model-assisted derotation is essential.  The scalar reference link supplies $5.893\times10^5$ phase tags in 100 s.  At 1 GHz this is one signal photon per $1.70\times10^5$ RF cycles.  For $a_{m,{\rm eff}}=0.5$--0.7 and negligible background, the accumulated phasor has ${\rm SNR}_{\rm AM}=192$--269.  With $\sigma_\phi\simeq{\rm SNR}_{\rm AM}^{-1}$, Eq.~(\ref{eq:shot}) gives $\sigma_\phi=3.72$--$5.21\,$mrad and $\sigma_{R,{\rm shot}}=88.8$--$124.3\,\mu$m.  The dedicated $\dot N_s=(3$--$5)\times10^4\,$s$^{-1}$ regime gives $30.5$--$55.1\,\mu$m in 100 s (phasor SNR 433--783 and phase error $1.28$--$2.31\,$mrad at the bounding $a_{m,{\rm eff}}$ values); alternatively, the scalar reference link needs about $0.88$--$1.72\,$ks to reach $30\,\mu$m. Equivalently, the $30\,\mu$m shot allocation at 1 GHz requires
\begin{equation}
 \frac{a_{m,{\rm eff}}^2\dot N_sT_{\rm int}}
 {1+\dot N_b/\dot N_s}\ge2.53\times10^6.
 \label{eq:photon_gate}
\end{equation}
For $T_{\rm int}=100\,$s, $a_{m,{\rm eff}}=0.7$, and $\dot N_b/\dot N_s=0.1$, this is $\dot N_s\ge5.68\times10^4\,$s$^{-1}$.  Thus the illustrative 2-m result $4.25\times10^4\,$s$^{-1}$ before $\eta_H$ does not by itself close the 100-s allocation; at 300 s the corresponding gate falls to $1.89\times10^4\,$s$^{-1}$.

Interleaved ranging among proposed distributed single HCCRs supplies
\begin{equation}
 \bm y_k=\{R_a,v_{r,a},\Delta R_{ab},\Delta v_{r,ab}\}_k,
 \label{eq:observables}
\end{equation}
with
\begin{equation}
 \bm C=\bm C_{\rm shot}+\bm C_{\rm atm}+\bm C_{\rm inst}
 +\bm C_{\rm osc}+\bm C_{\rm nl}+\bm C_H+\bm C_{\rm model} .
 \label{eq:C}
\end{equation}
The subscripts ${\rm shot}$, ${\rm atm}$, ${\rm inst}$, ${\rm osc}$, ${\rm nl}$, $H$, and ${\rm model}$ denote photon noise, residual atmosphere, internal optical--RF metrology, oscillator noise, amplitude-modulation-to-phase-modulation (AM-to-PM) conversion plus multi-tone nonlinearity, HCCR phase-center and geometry effects, and global lunar-model error, respectively.  A favorable dedicated-facility allocation is
\begin{equation}
 \sigma_R^2=(30^2+60^2+40^2+5^2+25^2+10^2)\,\mu{\rm m}^2,
 \label{eq:budget}
\end{equation}
for shot, atmosphere, instrument, oscillator, nonlinearity, and HCCR white terms, respectively.  The quadrature sum assumes independent contributions within one accepted window; measured cross-correlations and longer-timescale components instead enter the full matrix in Eq.~(\ref{eq:C}).  The sum gives $\sigma_R=82.8\,\mu$m, denoted an $80\,\mu$m-class requirement below.  The science curves use $80\,\mu$m exactly as a nominal target; substituting the literal allocation raises each unit-factor covariance-branch curve by $(82.8/80)^2=1.071$.  The budget is projected rather than demonstrated.  The $C_{\rm model}$ term in Eq.~(\ref{eq:C}) is deliberately absent from this per-window white allocation: model error is propagated as a correlated process and through fitted ephemeris nuisance parameters rather than assigned an independent white variance.  The non-shot terms in Eq.~(\ref{eq:budget}) have quadrature sum $\sigma_{R,{\rm nonshot}}=77.1\,\mu$m.  A $50\,\mu$m total uncertainty with a $30\,\mu$m photon term therefore requires the combined non-shot allocation to fall to at most $40\,\mu$m.  The $50\,\mu$m case is a coordinated system-level stretch; more photons alone cannot realize it.

\begin{table*}[t]
\caption{Reference AM-CW measurement chain and evidentiary status.  A
``derived'' entry follows from the stated model inputs; ``projected'' denotes
a facility or target performance assumption; and ``validation input'' is a
quantity that must be measured for the experiment-level likelihood.}
\label{tab:measurement_status}
\begin{ruledtabular}
\begin{tabular}{@{}p{0.22\textwidth}p{0.26\textwidth}
                    p{0.26\textwidth}p{0.22\textwidth}@{}}
Quantity & Reference value & Basis & Status\\
Transmitter/receiver & $P_0=1\,$kW, $\lambda=1064\,$nm,
$D_{\rm tx}=D_{\rm rx}=1\,$m & Refs.~\cite{TuryshevCW2025,TuryshevAMCW2025}
& projected design inputs\\
Uplink angle & $r_0=0.20\,$m, $\theta_{\rm tx}=5.476\,\mu$rad
& diffraction plus turbulence & projected good-seeing input\\
HCCR scalar reference & $D_H=0.10\,$m, $\theta_{\rm ret}=12.981\,\mu$rad
& circular-aperture first null & projected target input\\
End-to-end factor & $\eta_{\rm eff}=0.194208$
& stated component product & projected scalar input\\
HCCR transfer & $\eta_H(\lambda,\alpha_{\rm va},\iota,T_H)$
& vector far field and phase center & validation input\\
Reference signal rate & $\dot N_s=5.893\times10^3\,$s$^{-1}$
& Eq.~(\ref{eq:link}), $\eta_H=1$ & derived normalization\\
Reference 100-s photon term & $88.8$--$124.3\,\mu$m
& Eq.~(\ref{eq:shot}), $a_{m,{\rm eff}}=0.5$--0.7, negligible background
& derived projection\\
Dedicated 100-s photon gate & $\dot N_s\ge5.68\times10^4\,$s$^{-1}$
& Eq.~(\ref{eq:photon_gate}) & system requirement\\
Absolute range cases & $82.8\,\mu$m; $50\,\mu$m stretch
& Eq.~(\ref{eq:budget}); non-shot $\le40\,\mu$m & projected allocations\\
\end{tabular}
\end{ruledtabular}
\end{table*}

The transition from photon-limited to system-limited operation is explicit in this budget.  Since $\sigma_{R,{\rm shot}}^2\propto\dot N_s^{-1}$ and the covariance-branch threshold is proportional to $\sigma_R^2$, its local photon rate elasticity at the $30\,\mu$m allocation is
\begin{equation}
 {\cal E}_{N}\equiv-
 \frac{\partial\ln\Ogw^{95}}{\partial\ln\dot N_s}
 =\frac{\sigma_{R,{\rm shot}}^2}{\sigma_R^2}=0.131 .
 \label{eq:photon_elasticity}
\end{equation}
For the $82.8\,\mu$m reference allocation, holding every non-shot term fixed while taking the photon rate to infinity would reduce the threshold by only $1-\sigma_{R,{\rm nonshot}}^2/\sigma_R^2=13.1\%$, an improvement factor of 1.15.  For the $50\,\mu$m stretch allocation, the corresponding photon-rate elasticity is $(30/50)^2=0.36$ and the absolute flux-only improvement ceiling is $(50/40)^2=1.56$.  Beyond the dedicated photon gate, atmospheric, instrumental, nonlinear, and HCCR terms therefore determine the useful science gain from additional power.

Incoherent, unmodulated sky and lunar background has zero demodulated mean but contributes Poisson variance in Eq.~(\ref{eq:shot}).  Transmitter leakage, RF pickup, detector nonlinearity, or diffuse laser backscatter at a modulation tone can instead be coherent and must be bounded with off-target, off-range, and off-tone channels.  Spatial field stops, a Doppler-tracked narrow optical passband, cooled low-dark-count detection, off-tone background channels, and rejection of scintillating or high-background windows are all required.  At $\dot N_s=4\times10^4\,$s$^{-1}$, $a_{m,{\rm eff}}=0.6$, and $T_{\rm int}=100\,$s, requiring $\dot N_b/\dot N_s\le0.1$ keeps the photon term below $42\,\mu$m.  The $60\,\mu$m atmospheric entry is likewise a post-selection and post-calibration requirement, not an established absolute-LLR residual.  Its nightly and synoptic correlations must be propagated.  Dual-wavelength ranging can monitor dispersive delay but does not remove nearly nondispersive neutral-atmosphere piston, station motion, or hardware drift \cite{Abshire1980,MendesPavlis2004,He2023}. If the atmospheric residual were $100\,\mu$m rather than $60\,\mu$m, with the other allocations unchanged, Eq.~(\ref{eq:budget}) would give $\sigma_R=115.1\,\mu$m and would raise the unit-factor $80\,\mu$m SGWB thresholds by $(115.1/80)^2=2.07$.

There is no lunar clock: the HCCR is passive.  All frequency and phase references are terrestrial, and coherence is required within accepted $10^2$--$10^3\,$s windows, not continuously for five years.  The $80\,\mu$m target is $3.35\,$mrad at 1 GHz and $0.53\,$ps of two-way delay; the $40\,\mu$m instrument allocation is $1.68\,$mrad or $0.27\,$ps.  An uncompensated modulation scale would require $|\delta f_m/f_m|<2.1\times10^{-13}$ at $80\,\mu$m, so internal path and frequency references must be continuously monitored rather than merely assumed stable.  More specifically, the $5\,\mu$m oscillator allocation in Eq.~(\ref{eq:budget}) corresponds over the $2.56\,$s round trip to
\begin{equation}
\sigma_{R,{\rm osc}}\simeq\LEM\sigma_y(2.56\,{\rm s})
 =3.84\mum\left[\frac{\sigma_y(2.56\,{\rm s})}{10^{-14}}\right],
 \label{eq:oscillator_requirement}
\end{equation}
and therefore requires a residual fractional-frequency instability $\sigma_y(2.56\,{\rm s})\lesssim1.3\times10^{-14}$ after reference and path calibration, where $\sigma_y$ here denotes the round-trip sensitivity-weighted fractional-frequency average.  This is a low-frequency allocation scale, not a complete clock qualification: the verified budget must propagate the measured phase-noise spectrum through the two-way-delay and normal-point averaging transfer, because one Allan-deviation value does not bound all drift and phase noise. ``Continuous wave'' describes illumination inside each accepted window, not uninterrupted campaign operation.  For 100-s windows, an effective accepted-window cadence $\nuEff=500\,$yr$^{-1}$ is 13.9 effective h yr$^{-1}$ and requires $70$--$280\,$scheduled h yr$^{-1}$ at 5--20\% net efficiency; the $1500\,$yr$^{-1}$ higher-performance case requires 41.7 effective and $210$--$840\,$scheduled h yr$^{-1}$.

Thus the SGWB gain is not ``more photons'' alone.  It is an absolute common-mode range observable with a known time-domain response, accompanied by derivative and reflector-difference controls whose residual covariances are estimated rather than assumed away.

\emph{Response-calibrated reach.}--The local form of the accumulated broadband response makes the time law explicit.  Near $n=2$, the dominant phase-like range covariance has the schematic low-eccentricity form $C_{R,2}^{\rm GW}(t,t')={\cal A}_2^2\Ogw(\fii)K_2(t,t')$, with local shape kernel
\begin{align}
 K_2(t,t')&={\cal I}_{\rm IW}(t,t')
 \cos[2\pi\fii(t-t')],
 \label{eq:local_kernel}\\
 {\cal I}_{\rm IW}(t,t')&=\frac{t_<^2(3t_>-t_<)}{6}.
 \label{eq:iw_kernel}
\end{align}
where ${\cal I}_{\rm IW}$ is the integrated-Wiener kernel, $t$ and $t'$ are measured from the campaign start, $t_<=\min(t,t')$, and $t_>=\max(t,t')$; ${\cal A}_2$ is fixed by $B_{2,ij}$ and the range partials in Eq.~(\ref{eq:range_response}).  Thus $C_{R,2}^{\rm GW}(t,t)={\cal A}_2^2\Ogw t^3/3$: the SGWB drives a random walk in an orbital-rate variable, and range records the accumulated orbital phase.  For $N_{\rm eff}$ white, well-distributed range points, the weak-signal Fisher information for the local SGWB amplitude $A_{\rm GW}\equiv\Ogw(\fii)$ scales as $F_{A_{\rm GW}A_{\rm GW}}\propto N_{\rm eff}^2T_{\rm obs}^6/\sigma_R^4$, giving
\begin{equation}
 \Ogw^{95}\propto
 \frac{\sigma_R^2}{N_{\rm eff}T_{\rm obs}^3}
 =\frac{\sigma_R^2}{\nuEff T_{\rm obs}^4}.
 \label{eq:ideal_scaling}
\end{equation}
Equivalently, $h_c^{95}\propto T_{\rm obs}^{-2}$ at fixed cadence \cite{Hui2013,FosterMethods2025}.  Eq.~(\ref{eq:ideal_scaling}) applies to a broadband SGWB smooth across $\Delta f\sim T_{\rm obs}^{-1}$.  An exactly resonant unresolved monochromatic stochastic component instead has the ideal $\Ogw^{95}\propto T_{\rm obs}^{-5}$ law, and an off-resonance component scales as $T_{\rm obs}^{-3}$ \cite{FosterMethods2025}.

To set the response normalization independently of the AM-CW metrology model, we use the published Earth--Moon likelihood ordinate $\Omega_{\rm BJ}^{95}(\fii)=6.2\times10^{-6}$ as a conditional power-law-integrated (PI) envelope anchor \cite{BlasJenkinsPRL2022}.  It is not a universal monochromatic likelihood value.  That calculation represented $T_{\rm BJ}=15\,$yr of Apache Point Observatory Lunar Laser-ranging Operation (APOLLO) data by approximately $260$ normal points yr$^{-1}$ at $\sigma_{\rm BJ}=3\,$mm; the ``approximately 1000 measurements'' in the same paper was a separate back-of-the-envelope illustration and did not normalize the likelihood forecast.  The subscript ${\rm BJ}$ denotes this Blas--Jenkins reference calculation.

\emph{Sensitivity-factor notation.}--We reserve calligraphic $\mathcal D$ for dimensionless sensitivity-transfer or diagnostic quantities.  Let $N_{\rm BJ}\simeq3900$, $N_{\rm eff}\equiv\nuEff T_{\rm obs}$, and $r\equiv\sigma_R/\sigma_{\rm BJ}$, and define the unit-factor, pure-covariance anchor transfer
\begin{equation*}
 \Omega_0^{95}(\fii)\equiv\Omega_{\rm BJ}^{95}(\fii)r^2
 \frac{N_{\rm BJ}}{N_{\rm eff}}
 \left(\frac{T_{\rm BJ}}{T_{\rm obs}}\right)^3.
\end{equation*}
The net experiment transfer is
\begin{equation*}
 \Dtot\equiv\frac{(\Ogw^{95})_{\rm AM,full}}{\Omega_0^{95}}
 =\Dnorm\Dcamp\Dcov.
\end{equation*}
Thus $\Dtot$ is a sensitivity-threshold ratio, not necessarily a degradation factor: $\Dtot=1$ reproduces the unit-factor anchor forecast, smaller values improve the threshold, and larger values weaken it.  Only $\Dnorm$, $\Dcamp$, and $\Dcov$ enter this product.  The factor $\Dnorm$ corrects the reference normalization for its mean--covariance information mixture; $\Dcamp$ transfers the sampled AM-CW response, timestamps, observables, and spectral shape with nuisance parameters fixed; and $\Dcov$ measures the subsequent loss from colored noise, deterministic-parameter fitting, and covariance-hyperparameter marginalization.  The first two are algebraically linked bookkeeping factors, while $\Dcov$ is an experiment-level loss.  The quantities $\Did$ and $\Dmax$ introduced below are, respectively, a local identifiability diagnostic and a source-dependent allowable ceiling; neither is multiplied into $\Dtot$.  With these definitions, the response-calibrated threshold is
\begin{align}
 \Ogw^{95}(\fii)&=\Dtot\,\Omega_0^{95}(\fii)
 =\Dtot\,\Omega_{\rm BJ}^{95}
 \left(\frac{\sigma_R}{\sigma_{\rm BJ}}\right)^2\nonumber\\
 &\quad\times
 \left(\frac{N_{\rm BJ}}{N_{\rm eff}}\right)
 \left(\frac{T_{\rm BJ}}{T_{\rm obs}}\right)^3\nonumber\\
 &=1.86\times10^{-7}\Dtot
 \left(\frac{\sigma_R}{80\mum}\right)^2
 \left(\frac{500\,{\rm yr}^{-1}}{\nuEff}\right)\nonumber\\
 &\quad\times
 \left(\frac{5\,{\rm yr}}{T_{\rm obs}}\right)^4,
 \label{eq:campaign}
\end{align}
To evaluate the reference-normalization factor, write $F_{\rm BJ}=K_m/\sigma_{\rm BJ}^2+K_c/\sigma_{\rm BJ}^4$ with $K_m,K_c\ge0$ and define its covariance-information fraction $w_{\rm cov,BJ}$.  Relative to the pure-covariance $r^2$ transfer used in Eq.~(\ref{eq:campaign}),
\begin{align}
 w_{\rm cov,BJ}&\equiv
 \frac{K_c/\sigma_{\rm BJ}^4}{F_{\rm BJ}},\nonumber\\
 \Dnorm(r,w_{\rm cov,BJ})
 &= [w_{\rm cov,BJ}+(1-w_{\rm cov,BJ})r^2]^{-1/2},\nonumber\\
 1&\le\Dnorm\le r^{-1}\qquad(0<r\le1).
 \label{eq:dnorm}
\end{align}
The lower endpoint is a covariance-dominated reference likelihood; the upper is mean dominated.  At $80\,\mu$m, $\Dnorm=1$--$37.5$; at $50\,\mu$m, it is $1$--$60$.  Larger $\Dnorm$ weakens the nominal $r^2$ improvement and enters the sensitivity only through $\Dtot$.  This is a fixed-reference normalization-stress envelope, not a small error bar or an independent bound on $\Dtot$; $\Dnorm$ and $\Dcamp$ are linked by Eq.~(\ref{eq:dcamp}).

The response mixture is directly diagnosable by controlled likelihood reruns. Holding the timestamps and response fixed, define
\begin{equation}
 p_\sigma(\sigma_R)\equiv
 \frac{d\ln\Ogw^{95}}{d\ln\sigma_R}=1+w_{\rm cov}(\sigma_R),
 \label{eq:precision_exponent}
\end{equation}
where the equality follows from $F=K_m\sigma_R^{-2}+K_c\sigma_R^{-4}$ and $w_{\rm cov}(\sigma_R)\equiv K_c\sigma_R^{-4}/F(\sigma_R)$; hence $w_{\rm cov}(\sigma_{\rm BJ})=w_{\rm cov,BJ}$.  Thus $p_\sigma=1$ and 2 are the pure-mean and pure-covariance limits.  Rerunning the same sampled likelihood at neighboring assigned white-noise levels measures the mixture rather than selecting it by assumption.  For the five-year $50\,\mu$m, $1500\,\mathrm{yr}^{-1}$ compact-binary requirement with $\Dcamp=\Dcov=1$, the reference mixture must satisfy $w_{\rm cov,BJ}\ge0.759$, equivalently $p_\sigma(\sigma_{\rm BJ})\ge1.759$.

The envelope is larger than percent-level changes in the optical budget. Holding $\Dcamp=\Dcov=1$ only to expose its numerical leverage, the five-year $80\,\mu$m, $500\,\mathrm{yr}^{-1}$ threshold spans $1.86\times10^{-7}$--$6.96\times10^{-6}$ across the allowed $\Dnorm$ endpoints.  For the five-year $50\,\mu$m, $1500\,\mathrm{yr}^{-1}$ case the corresponding interval is $2.42\times10^{-8}$--$1.45\times10^{-6}$.  These endpoint calculations do not treat $\Dnorm$ and $\Dcamp$ as independent physical errors; they show that benchmark-source reach requires the reference likelihood to contain a large covariance-information fraction or the sampled response to provide a compensating enhancement.

For normalized local spectral shape $s_2(f)\equiv\Ogw(f)/\Ogw(\fii)$, the actual AM-CW sampling and full response kernel are separated through
\begin{align}
 \Dcamp[s_2,\{t_k\};r,w_{\rm cov,BJ}]
 &\equiv\frac{(\Ogw^{95})_{\rm AM,white,fixed}}
 {\Omega_{\rm BJ}^{95}r^2}\nonumber\\
 &\quad\times\frac{N_{\rm eff}}{N_{\rm BJ}}
 \left(\frac{T_{\rm obs}}{T_{\rm BJ}}\right)^3\Dnorm^{-1}.
 \label{eq:dcamp}
\end{align}
The campaign-transfer factor $\Dcamp$ has no a priori inequality about unity: smaller values represent a response enhancement and larger values a weaker sampled response.  It is computed with the actual timestamps, AM-CW observables, fixed physical response, and likelihood statistic, integrating both the drift and covariance responses in Eqs.~(\ref{eq:range_mean}) and (\ref{eq:range_kernel}); information loss from fitting or marginalizing nuisance parameters belongs exclusively to $\Dcov$.  The factor $\Dcamp$ also carries the difference between the PI spectral family used for the anchor and the particular local shape $s_2(f)$; consequently the illustrative phase-transition and compact-binary spectra below cannot inherit exactly the same ordinate without integrating their shapes through ${\cal K}_{kl}$. Finally, we define the loss from colored residuals and fitted nuisance parameters on those same timestamps by
\begin{equation}
 \Dcov\equiv
 \frac{(\Ogw^{95})_{\rm AM,full}}
 {(\Ogw^{95})_{\rm AM,white,fixed}}\ge1.
 \label{eq:dcovdef}
\end{equation}
The denominator is, by definition, a white, fixed-parameter information upper bound for the same sampled problem, chosen to be no noisier in the signal subspace.  Thus larger $\Dcov$ is worse, and in the weak-signal Fisher limit $\Dcov\simeq(F_{AA}^{\rm white,fixed}/F_{AA}^{\rm full,marg})^{1/2}$.  This three-factor decomposition prevents response calibration, schedule, and noise marginalization from being hidden in one number or counted twice.  Experiment-level values of $\Dcamp$ and $\Dcov$ are obtained from realized timestamps, measured HCCR transfer functions, physical ephemeris partials, and colored-noise covariance; the unit-factor curves below specify the performance requirements those data products must satisfy.

At fixed duration, dense pair-rich covariance information gives the $N_{\rm eff}^{-1}$ threshold in Eq.~(\ref{eq:campaign}), whereas independent covariance blocks give only $N_{\rm eff}^{-1/2}$.  We write this operationally as
\begin{equation}
 \Ogw^{95}(\nuEff)=\Ogw^{95}(\nu_0)
 \left(\frac{\nu_0}{\nuEff}\right)^{\alpha_N},\qquad
 \frac12\le\alpha_N\le1,
 \label{eq:cadence_scaling}
\end{equation}
and do not equate raw windows with $N_{\rm eff}$ once atmospheric, geometry, or model correlations repeat.  The interval $1/2\le\alpha_N\le1$ describes statistics-limited regimes only.  Once a red systematic or model-degeneracy floor fixes the number of informative modes, the effective exponent tends toward zero and additional optical power or raw windows no longer improve the SGWB threshold.

A timestamped forecast can test these asymptotes directly through the nested sensitivity exponents
\begin{align}
 p_T&=-\frac{\ln[\Ogw^{95}(T_2)/\Ogw^{95}(T_1)]}
 {\ln(T_2/T_1)},\nonumber\\
 p_\nu&=-\frac{\ln[\Ogw^{95}(\nu_2)/\Ogw^{95}(\nu_1)]}
 {\ln(\nu_2/\nu_1)} .
 \label{eq:sensitivity_exponents}
\end{align}
The statistics-limited model gives $p_\nu=\alpha_N$ and $p_T=3+\alpha_N$, hence $p_T=3.5$--4.  Equivalently, only 17--21\% of the ideal five-year Fisher information is present after four years.  A five-year source crossing is accepted as converged only when nested-duration, cadence-thinning, and integration-window reruns recover stable exponents rather than a prior- or endpoint-driven change.

The published reference uses the likelihood-ratio threshold $\Lambda=3.841$, the conventional 95\% $\chi_1^2$ critical value adopted in that calculation.  A convenient source-planning conversion is therefore
\begin{equation}
 \Ogw^{\rm nom,5\sigma}=\zeta_5\Ogw^{95},\qquad
 \zeta_5=\frac{5}{\sqrt{3.841}}=2.55 .
 \label{eq:rho5}
\end{equation}
Because $\Ogw\ge0$ is a boundary parameter and a real search scans spectral shapes, this is not a calibrated false-alarm or detection-power threshold; those require null and injection simulations of the final likelihood.

The sampling window is
\begin{equation}
 {\cal W}(f)=\sum_kw_k e^{-2\pi ift_k}.
 \label{eq:window}
\end{equation}
Here $w_k$ is the analysis weight assigned to the accepted measurement at $t_k$, with $w_k=1$ for an unweighted sampling window.  A reproducible local identifiability screen follows directly from Eq.~(\ref{eq:local_kernel}). Specifically, $(\bm K_2)_{kl}=K_2(t_k,t_l)$ on the sampled timestamps.  The common overall factor ${\cal A}_2^2\Ogw(\fii)$ is omitted because it cancels from $q_{\cal S}$.  Let $\bm\Pi_{\cal S}$ remove a nuisance set ${\cal S}$ from white data.  The retained covariance information and corresponding line-fit loss are
\begin{equation}
 \begin{aligned}
 q_{\cal S}&=\frac{{\rm Tr}[(\bm\Pi_{\cal S}\bm K_2
 \bm\Pi_{\cal S})^2]}{{\rm Tr}(\bm K_2^2)}
 =\frac{F_{AA}^{\rm local,proj({\cal S})}}
 {F_{AA}^{\rm local,unproj}},\\
 \Did&=q_{\cal S}^{-1/2}.
 \end{aligned}
 \label{eq:identifiability_screen}
\end{equation}
For midpoint sampling $t_k=(k+1/2)/\nuEff$, $k=0,\ldots,N_{\rm eff}-1$, at $500\,\mathrm{yr}^{-1}$, we use $\bm\Pi_{\cal S}=\bm I-\bm X_{\cal S} (\bm X_{\cal S}^{T}\bm X_{\cal S})^{-1}\bm X_{\cal S}^{T}$.  The base set $\{1,t-\bar t,(t-\bar t)^2\}$ gives $\Did=1.00005$ at five years.  Adding free sine/cosine pairs at the half-draconic or half-anomalistic frequencies gives 1.605 or 1.095; fitting both gives 1.723. A free pair exactly at $f_2$ gives 2.530.  The five-year, both-sideband value changes from 1.723280 to 1.723274 as the midpoint grid is refined from 250 to $1000\,\mathrm{yr}^{-1}$.  By construction $\Did\ge1$ and larger values indicate greater loss in this simplified white-noise local screen.  It is diagnostic only and is not multiplied into $\Dtot$.  In the full likelihood the corresponding deterministic-fit loss is already contained in $\Dcov$ through $q_{\rm det}$, so multiplying $\Did$ by $\Dcov$ would double count it.  Only when the full analysis uses the same white covariance, timestamps, kernel, and nuisance subspace without priors or hyperparameter marginalization does $\Did=q_{\rm det}^{-1/2}$.

\begin{figure*}[t]
\includegraphics[width=\textwidth]{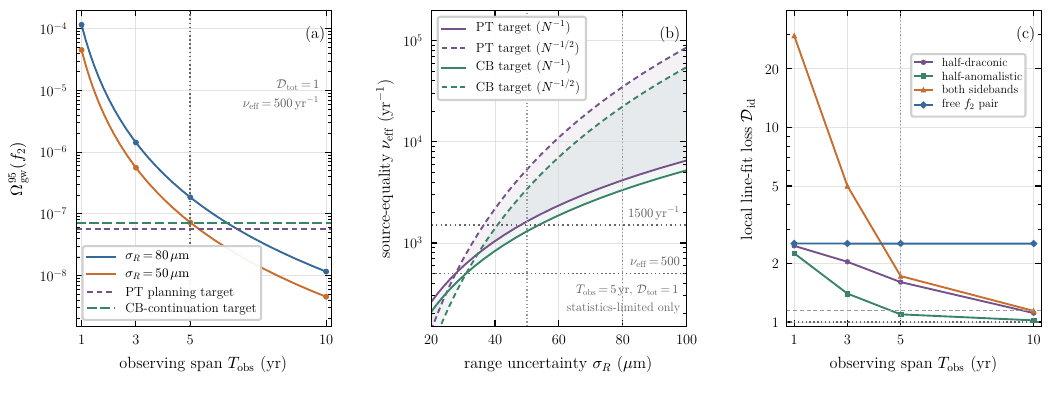}
\caption{AM-CW SGWB requirements at $\fii=0.847245\,\uHz$.
(a) Conditional response-anchored 95\% thresholds for
$\nuEff=500\,$yr$^{-1}$ and $\Dtot=1$, following the broadband
$T_{\rm obs}^{-4}$ law; circles mark 1, 3, 5, and 10 yr.  PT denotes the
phase-transition planning target and CB the compact-binary-continuation
target.  Their intersections with the curves are 95\% upper-limit equalities,
not the $\zeta_5$ source-statistic equalities in panel (b).  (b) Effective
cadence required for five-year source-statistic equality.  Solid curves use
the pair-rich $N_{\rm eff}^{-1}$ limit and dashed curves the independent-block $N_{\rm eff}^{-1/2}$ limit; light shading spans statistics-limited exponents only.  The vertical guides mark the 50 and $80\,\mu$m design cases.  A correlated systematic floor can place the requirement above the shaded band or eliminate cadence gain.  A factor $\Dtot$ raises the required cadence by $\Dtot^{1/\alpha_N}$ at fixed precision when $\Dtot$ is held fixed as cadence changes.  (c) Local line-fit loss from Eq.~(\ref{eq:identifiability_screen}) for uniform midpoint sampling at $500\,\mathrm{yr}^{-1}$ after removing a quadratic trend and the indicated unconstrained quadrature pairs.  The horizontal guide marks the $\Dtot=1.148$ compact-binary allowance for the aggressive five-year case.  The discretization-converged kernel screen quantifies $\Did$ locally; it is diagnostic only, is not multiplied into $\Dtot$, and is not a global-ephemeris value of $\Dcov$.  All panels show calculated unit-factor
requirements or diagnostics, not observed limits.}
\label{fig:reach}
\end{figure*}

Five years gives $T_{\rm obs}^{-1}=6.34\,$nHz.  The half-draconic line at $f_{2,{\rm dr}}=0.850653\,\mu$Hz lies only $3.41\,$nHz from $f_2$ (a 9.30-year beat), while the half-anomalistic line at $f_{2,{\rm an}}=0.840084\,\mu$Hz is $7.16\,$nHz away.  Thus a five-year Fourier bin does not by itself separate all lunar sidebands; the time-domain ephemeris fit and its priors are a physical part of $\Dcov$. Table~\ref{tab:reach} lists the principal unit-factor requirements, and Fig.~\ref{fig:reach} shows their duration and cadence dependence.

\begin{table}[t]
\caption{Response-anchored AM-CW requirement cases at
$\fii=0.847245\,\uHz$.  Values multiply by the full-to-unit threshold ratio $\Dtot$ and use the pair-rich
$\alpha_N=1$ cadence law.  The last column is the nominal source-statistic
mapping in Eq.~(\ref{eq:rho5}), not a calibrated discovery threshold.}
\label{tab:reach}
\begin{ruledtabular}
\begin{tabular}{lccc}
Case & $\sigma_R$ & $\Ogw^{95}/\Dtot$ &
$\zeta_5\Ogw^{95}/\Dtot$\\
5 yr, $500\,$yr$^{-1}$ & $80\mum$ & $1.86\times10^{-7}$ & $4.74\times10^{-7}$\\
5 yr, $500\,$yr$^{-1}$ & $50\mum$ & $7.25\times10^{-8}$ & $1.85\times10^{-7}$\\
5 yr, $1500\,$yr$^{-1}$ & $50\mum$ & $2.42\times10^{-8}$ & $6.17\times10^{-8}$\\
10 yr, $500\,$yr$^{-1}$ & $80\mum$ & $1.16\times10^{-8}$ & $2.96\times10^{-8}$\\
10 yr, $500\,$yr$^{-1}$ & $50\mum$ & $4.53\times10^{-9}$ & $1.16\times10^{-8}$\\
\end{tabular}
\end{ruledtabular}
\end{table}

\emph{Illustrative source scales.}--The following spectra are not priors or predictions for this experiment; they are target amplitudes for the requirements calculation.  For sound-wave-dominated first-order phase transitions \cite{Caprini2016,Caprini2020,BlasJenkinsPRL2022},
\begin{align}
 f_{\rm sw}&\simeq19\,\uHz
 \left(\frac{T_*}{100\,{\rm GeV}}\right)
 \left(\frac{\beta/H_*}{v_w}\right)
 \left(\frac{g_*}{106.75}\right)^{1/6},\label{eq:fsw}\\
 \Ogw^{\rm peak}&\simeq5.7\times10^{-6}
 \left(\frac{v_w}{\beta/H_*}\right)
 \left(\frac{\kappa_{\rm sw}\alpha_{\rm PT}}{1+\alpha_{\rm PT}}\right)^2\nonumber\\
 &\quad\times
 \left(\frac{g_*}{106.75}\right)^{-1/3}S_{\rm sw}. \label{eq:PT}
\end{align}
Here $T_*$ is the transition temperature, $\beta^{-1}$ the transition duration, $H_*$ the Hubble rate at the transition epoch, $v_w$ the bubble-wall speed, $g_*$ the number of relativistic degrees of freedom, $\alpha_{\rm PT}$ the released vacuum energy in units of the radiation density, $\kappa_{\rm sw}$ the efficiency for acoustic bulk motion, and $S_{\rm sw}$ the finite-lifetime/spectral factor. Matching $f_{\rm sw}$ to $\fii$ gives
\begin{equation}
 T_*\simeq4.46\,{\rm GeV}
 \left(\frac{v_w}{\beta/H_*}\right)
 \left(\frac{106.75}{g_*}\right)^{1/6}.
 \label{eq:Tstar}
\end{equation}
Thus the lunar resonance selects GeV-to-sub-GeV transitions for $\beta/H_*\sim1$--10.  An intentionally optimistic phenomenological point, $\alpha_{\rm PT}=3$, $\kappa_{\rm sw}=0.7$, $v_w=1$, $\beta/H_*=10$, $S_{\rm sw}=0.3$, and $g_*=61.75$, gives $T_*=0.49\,$GeV and $\Ogw^{\rm peak}=5.66\times10^{-8}$.  It is not a prediction of Standard-Model quantum chromodynamics (QCD).

A compact-binary continuation with $\hc=A_{\rm PTA}(f/f_{\rm yr})^{-2/3}$, $f_{\rm yr}=1\,$yr$^{-1}$, and $A_{\rm PTA}=2.4\times10^{-15}$ gives $\hc(\fii)=2.68\times10^{-16}$ and $\Ogw(\fii)=7.08\times10^{-8}$ \cite{NANOGrav2023,NANOGravSMBHB2023,EPTA2023}.  This power-law continuation may bend before reaching $\mu$Hz frequencies, and a sparse compact-binary population need not be Gaussian within a bandwidth $1/T_{\rm obs}$; in that case a deterministic or non-Gaussian search replaces the SGWB covariance likelihood.  These two curves therefore remain planning benchmarks only.

Equations~(\ref{eq:campaign}) and (\ref{eq:rho5}) give the nominal margin
\begin{align}
 \Dmax&\equiv\frac{\Ogw^{\rm src}}{\zeta_5\Omega_0^{95}}
 =\frac{\Ogw^{\rm src}}{4.74\times10^{-7}}
 \left(\frac{80\mum}{\sigma_R}\right)^2\nonumber\\
 &\quad\times
 \left(\frac{\nuEff}{500\,{\rm yr}^{-1}}\right)
 \left(\frac{T_{\rm obs}}{5\,{\rm yr}}\right)^4 .
 \label{eq:dtotmax}
\end{align}
The quantity $\Dmax$ is a source-dependent allowable ceiling, not another factor in the sensitivity product.  Larger values provide more design margin, and nominal source-statistic equality requires $\Dtot\le\Dmax$.
At five years and $500\,$yr$^{-1}$, nominal source-statistic equality requires $\Dtot\le0.12$ and 0.15 for the phase-transition and compact-binary targets at $80\,\mu$m, and $\Dtot\le0.31$ and 0.38 at $50\,\mu$m.  Even at $\Dnorm=\Dcov=1$, these cases require $\Dcamp$ to supply the corresponding response enhancement relative to the reference transfer.  They define upper-limit design cases rather than benchmark detections.

For the higher-performance five-year $50\,\mu$m, $1500\,\mathrm{yr}^{-1}$ campaign, equality requires $\Dtot\le0.92$ for the phase-transition point and $\Dtot\le1.15$ for the compact-binary continuation.  Because $\Dnorm,\Dcov\ge1$, the former specifically requires $\Dcamp<0.92$, whereas the latter permits only a 15\% total-factor margin. The unrounded compact-binary allowance is $\Dmax=1.1477$.  If $\Dnorm=\Dcamp=1$, the compact-binary case must retain $F_{AA}^{\rm full,marg}/F_{AA}^{\rm white,fixed}\ge0.759$.  This future-campaign information-retention fraction is distinct from $w_{\rm cov,BJ}$, which measures the covariance contribution within the published reference likelihood; their identical rounded values under this particular stress test are coincidental.  The freely fitted two-sideband local screen in Eq.~(\ref{eq:identifiability_screen}) retains only 0.337, so physical ephemeris constraints, geometry telemetry, or auxiliary range channels are quantitatively load-bearing.  At $50\,\mu$m the exact pair-rich cadence requirements are $1.63\times10^3$ and $1.31\times10^3\,\mathrm{yr}^{-1}$ for the phase-transition and compact-binary targets; in the independent-block limit they are $5.35\times10^3$ and $3.42\times10^3\,\mathrm{yr}^{-1}$.  The ten-year $80\,\mu$m product margins are $1.91$ and $2.39$; under the same unit-factor stress convention these correspond to $w_{\rm cov,BJ}\ge0.274$ and $0.175$.  These are conditional diagnostics, not measurements of $w_{\rm cov,BJ}$, because $\Dnorm$ and $\Dcamp$ are linked by Eq.~(\ref{eq:dcamp}).

Only in the covariance-dominated broadband regime, with $\Dtot$ approximately constant over the interval considered, does a factor $\Dtot$ multiply a crossing time by $\Dtot^{1/4}$.  In that restricted regime, for $80\,\mu$m and $500\,$yr$^{-1}$ the unit-factor compact-binary and phase-transition crossings are $8.04$ and $8.50\,$yr, and a constant $\Dtot=5$ moves them to $12.0$ and $12.7\,$yr.  For $50\,\mu$m and $1500\,$yr$^{-1}$ the unit-factor crossings are $4.83$ and $5.11\,$yr.  The pure-mean endpoint generally has a different duration law and must not be extrapolated with $T_{\rm obs}^{-4}$; that change is carried by $\Dcamp(T_{\rm obs})$.  A five-year discovery is therefore an explicit commissioning target whose feasibility is decided by the measured response mixture and full AM-CW likelihood, not by extrapolating the range precision alone.

\emph{Absolute and differential modes.}--The absolute range from station $s$ to reflector $a$ can be written schematically as
\begin{align}
 R_{sa}(t)&=R_{EM}(t)
 +\hat{\bm n}^{T}\bm Q_{M\to R}(t)\bm x_a^M
 -\hat{\bm n}^{T}\bm Q_{E\to R}(t)\bm x_s^E\nonumber\\
 &\quad+R_{\rm atm}^{(s)}(t)+R_{\rm inst}^{(s)}(t)
 +\delta R_{H,a}(t)+\xi_{sa}(t).
 \label{eq:absolute}
\end{align}
Here $R_{EM}$ is the Earth--Moon center-to-center range contribution, $\hat{\bm n}$ is the line-of-sight unit vector, $\bm x_a^M$ and $\bm x_s^E$ are reflector and station coordinates in their body frames, and $\bm Q_{M\to R}$ and $\bm Q_{E\to R}$ rotate them into the ranging frame. The superscripts $M$ and $E$ denote Moon-fixed and Earth-fixed body frames, while $R$ denotes the ranging frame. The quantities $R_{\rm atm}^{(s)}$ and $R_{\rm inst}^{(s)}$ are atmospheric and instrumental path terms, and $\delta R_{H,a}$ is the HCCR phase-center/support contribution corresponding to $\delta\phi_{H,a}$ in Eq.~(\ref{eq:phase}).  The term $\xi_{sa}$ is the remaining zero-mean measurement residual for station $s$ and reflector $a$.

The SGWB signal is primarily in the common Earth--Moon orbital term and in the osculating elements that determine it. A pure differential range between CCRs separated by lunar baseline $B$ cancels that common term to first order,
\begin{align}
 \Delta R_{s,ab}(t)&\simeq
 \hat{\bm n}^{T}\bm Q_{M\to R}(t)(\bm x_a^M-\bm x_b^M)\nonumber\\
 &\quad+{\cal O}\!\left(\frac{B}{\LEM}\delta R_{EM}^{\rm GW}\right)
 +\Delta\xi_{s,ab}(t) .
 \label{eq:diffobs}
\end{align}
Here $\Delta\xi_{s,ab}(t)\equiv\xi_{sa}(t)-\xi_{sb}(t)$ is the corresponding differential measurement residual. For $B=10^3\,{\rm km}$ and $\sigma_{\Delta R}=20$--$50\mum$,
\begin{equation}
 h_{\Delta,{\rm eq}}\equiv\frac{2\sigma_{\Delta R}}{B}
 =(4\mbox{--}10)\times10^{-11},
 \label{eq:hDelta}
\end{equation}
where $h_{\Delta,{\rm eq}}$ is not an estimate of $h_c$ or $\Ogw$.  It is only an equivalent direct-baseline strain-noise scale obtained from the maximal geometric response $|\delta(\Delta R)|=hB/2$.  Equating that scale to $2\sigma_{\rm eff}^{\Delta}/\LEM$ gives
\begin{equation}
 \sigma_{\rm eff}^{\Delta}=\sigma_{\Delta R}\frac{\LEM}{B}
 =7.7\mbox{--}19\,{\rm mm}.
 \label{eq:effdiff}
\end{equation}
Inserted only as a scalar comparison in Eq.~(\ref{eq:campaign}), it gives $\Ogw^{\Delta,95}\simeq1.7\times10^{-3}$--$1.1\times10^{-2}$ for five years and $500\,$yr$^{-1}$.  Differential LLR is therefore not the SGWB amplitude channel.  It constrains reflector coordinates, librations, local tides, differential atmosphere, and target/instrument terms that otherwise project into the absolute range fit \cite{Zhang2022,Zhang2024}.

Here $v_r=dR/dt$ is the one-way line-of-sight range rate.  AM-CW phase slope does not measure transverse velocity directly; transverse state components enter through the changing line of sight and global dynamics.  Range and range rate estimated from the same phase window are correlated.  If the local phase design matrix $\bm X_k$ has columns $1$ and $t-\bar t$, and $\bm C_{\phi,k}$ is the covariance of the phase samples within window $k$, their white-noise covariance is
\begin{equation}
 \bm C_k^{(R,v_r)}=\left(\frac{c}{4\pi f_m}\right)^2
 (\bm X_k^T\bm C_{\phi,k}^{-1}\bm X_k)^{-1}.
 \label{eq:range_rate_cov}
\end{equation}
For a centered uniform window of duration $T_{\rm int}$ this gives $\sigma_{v_r,{\rm shot}}\simeq \sqrt{12}\sigma_{R,{\rm shot}}/T_{\rm int}$.  Thus $0.3\mums$ is a favorable several-hundred-second phase-slope allocation, not an independent consequence of the $80\,\mu$m total range budget \cite{Williams2018,TuryshevAMCW2025}.

As a transparent carrier-slope diagnostic, consider an uncorrelated sinusoidal range component at $\fii$.  Then
\begin{equation}
 \frac{\sigma_{v_r}}{2\pi\fii}=5.6\,{\rm cm}
 \left(\frac{\sigma_{v_r}}{0.3\mums}\right),
 \label{eq:vrf2}
\end{equation}
and its direct white-noise information relative to range is only
\begin{align}
 \frac{F_{v_r}}{F_R}\simeq
 \left(\frac{2\pi\fii\sigma_R}{\sigma_{v_r}}\right)^2
 &=2.0\times10^{-6}
 \left(\frac{\sigma_R}{80\mum}\right)^2
 \nonumber\\
 &\quad\times
 \left(\frac{0.3\mums}{\sigma_{v_r}}\right)^2.
 \label{eq:rate_direct_gain}
\end{align}
This $2.0\times10^{-6}$ ratio applies to that resolved sinusoidal carrier; the accumulated stochastic envelope also contributes derivatives that must be evaluated with the joint kernel below.  Equation~(\ref{eq:rate_direct_gain}) nevertheless shows why a large direct low-frequency gain should not be assumed.  The exact gain, including nuisance and alias separation, must be computed rather than asserted.  We define
\begin{equation}
 G_v\equiv
 \left(\frac{F_{AA,R+v_r}^{\rm marg}}
 {F_{AA,R}^{\rm marg}}\right)^{1/2}
 =\frac{\sigma_{A,R}}{\sigma_{A,R+v_r}}.
 \label{eq:range_rate_gain}
\end{equation}
No universal value is assigned to $G_v$: it depends on the actual timestamps, the within-window $R$--$v_r$ covariance, and the nuisance design matrix.  The same $0.3\mums$ gives sinusoidal amplitude-equivalent range scales of $4.1\,$mm, $0.17\,$mm, and $48\,\mu$m at periods of one day, one hour, and $10^3\,$s, respectively, illustrating why it is useful for rejecting short-period station, tide, libration, calibration, and phase-slip modes even though the carrier diagnostic in Eq.~(\ref{eq:rate_direct_gain}) is small.

\emph{Detection analysis.}--Let $A_{\rm GW}\equiv\Ogw(\fii)$, reserving $\Omega$ for the orbital node.  At the actual observation times the model is
\begin{align}
 \mu_k(A_{\rm GW},\bm q)&=h(t_k,\bm q)
 +A_{\rm GW}m_{{\rm GW},k},\nonumber\\
 y_k&=\mu_k(A_{\rm GW},\bm q)+\xi_k,\qquad
 H_{ka}=\frac{\partial h(t_k,\bm q)}{\partial q_a},
 \label{eq:observation_model}
\end{align}
with covariance
\begin{equation}
 \bm C_y(A_{\rm GW},\bm\eta)=\bm C_0(\bm\eta)
 +A_{\rm GW}\bm K_{\rm GW}.
 \label{eq:Cy}
\end{equation}
Here $\xi_k$ is a zero-mean measurement residual with the covariance specified below.  The vector $\bm q$ contains deterministic station, ephemeris, reflector, geophysical, and lunar-orientation parameters; $\bm\eta$ contains covariance hyperparameters.  Below, the Fisher subscript $A$ denotes $A_{\rm GW}$, not the orbital node $\Omega$.  For the local spectral shape defined above,
\begin{align}
 m_{{\rm GW},k}&=\int d\ln f\,s_2(f){\cal M}_k(f),\nonumber\\
 K_{{\rm GW},kl}=\int s_2(f){\cal K}_{kl}(f)\,d\ln f,
 \label{eq:bandkernel}
\end{align}
evaluated on the same timestamps and two-way light-time model as the data. In the short-window derivative approximation, the joint range--range-rate mean and covariance blocks are
\begin{align}
 \bm m_{\rm GW}^{(R,v)}(t)&=
 \begin{pmatrix}m_{\rm GW}(t)\\ \partial_t m_{\rm GW}(t)\end{pmatrix},\nonumber\\
 \bm K_{\rm GW}^{(R,v)}(t,t')&=
 \begin{pmatrix}
 K_{\rm GW}&\partial_{t'}K_{\rm GW}\\
 \partial_tK_{\rm GW}&\partial_t\partial_{t'}K_{\rm GW}
 \end{pmatrix}.
 \label{eq:joint_kernel}
\end{align}
The exact sampled blocks apply the same finite-window phase-slope functional used in Eq.~(\ref{eq:range_rate_cov}) to $m_{\rm GW}$ and to each argument of $K$; the derivatives above are accurate when the window is short compared with a $\mu$Hz period.

At the null covariance $\bm C_0$, linearized mean-parameter fitting is represented by the generalized residual projector
\begin{equation}
 \bm\Pi_0=\bm C_0^{-1}-\bm C_0^{-1}\bm H
 (\bm H^T\bm C_0^{-1}\bm H)^{-1}\bm H^T\bm C_0^{-1}.
 \label{eq:projector}
\end{equation}
Equation~(\ref{eq:projector}) is shown for a full-rank nuisance basis.  A production analysis must use an explicitly constrained basis, a generalized inverse, or add prior precision to the central normal matrix as appropriate. In the residual-contrast (restricted-likelihood) construction, the projected weak-signal information contains both the noise-induced drift and covariance channels,
\begin{equation}
 F_{AA}^{\rm proj}=\bm m_{\rm GW}^T\bm\Pi_0\bm m_{\rm GW}
 +\frac12{\rm Tr}(\bm\Pi_0\bm K_{\rm GW}\bm\Pi_0\bm K_{\rm GW}).
 \label{eq:Fisher}
\end{equation}
For mean and covariance derivatives $\bm\mu_{,A}$ and $\bm C_{,A}$ with $A\in\{A_{\rm GW},\eta_a\}$, define
\begin{equation}
 F_{AB}=\bm\mu_{,A}^T\bm\Pi_0\bm\mu_{,B}
 +\frac12{\rm Tr}(\bm\Pi_0\bm C_{,A}\bm\Pi_0\bm C_{,B}),
 \label{eq:covFisher}
\end{equation}
and marginalize covariance hyperparameters with
\begin{equation}
 F_{AA}^{\rm marg}=F_{AA}^{\rm proj}
 -F_{Aa}(F_{ab})^{-1}F_{bA},\qquad
 \sigma_A^2=(F_{AA}^{\rm marg})^{-1}.
 \label{eq:margF}
\end{equation}
This distinguishes deterministic mean fitting, which requires Eq.~(\ref{eq:projector}), from covariance-hyperparameter marginalization, for which the Schur complement is appropriate.

The total covariance loss can be reported as three information contractions,
\begin{align}
 q_{\rm noise}&=\frac{F_{AA}^{\rm full,fixed}}
 {F_{AA}^{\rm white,fixed}},&
 q_{\rm det}&=\frac{F_{AA}^{\rm full,proj}}
 {F_{AA}^{\rm full,fixed}},\nonumber\\
 q_{\rm hyp}&=\frac{F_{AA}^{\rm full,marg}}
 {F_{AA}^{\rm full,proj}},&
 \Dcov&\simeq(q_{\rm noise}q_{\rm det}q_{\rm hyp})^{-1/2}.
 \label{eq:information_contractions}
\end{align}
Here $q_{\rm noise}$ measures colored-noise loss, $q_{\rm det}$ contraction from deterministic-parameter fitting, and $q_{\rm hyp}$ the additional contraction from covariance-hyperparameter marginalization.  The latter has the exact Fisher multiple-correlation diagnostic
\begin{equation}
 {\cal R}_{A|\eta}^{\,2}\equiv
 \frac{F_{Aa}(F_{ab})^{-1}F_{bA}}{F_{AA}^{\rm proj}}
 =1-q_{\rm hyp}.
 \label{eq:multiple_correlation}
\end{equation}
The experiment-level analysis also reports the signal--parameter correlations from the inverse prior-augmented information matrix, its effective rank and eigenvalue spectrum, and the change in each $q$ under prior relaxation.

For null residuals $\bm r=\bm y-\bm h(\widehat{\bm q})$, the joint mean--covariance score is
\begin{equation}
 U_A=\bm\mu_{,A}^T\bm\Pi_0\bm r
 +\frac12[\bm r^T\bm\Pi_0\bm C_{,A}\bm\Pi_0\bm r
 -{\rm Tr}(\bm\Pi_0\bm C_{,A})].
 \label{eq:score}
\end{equation}
After eliminating covariance nuisances,
\begin{equation}
 U_A^{\rm eff}=U_A-F_{Aa}(F_{ab})^{-1}U_b,\qquad
 \widehat A_{\rm GW}^{\rm unc}=(F_{AA}^{\rm marg})^{-1}U_A^{\rm eff}.
 \label{eq:estimator}
\end{equation}
This is the unconstrained, locally optimal linear-plus-quadratic one-step estimator in the weak-signal limit.  The physical fit imposes $A_{\rm GW}\ge0$; null and injection ensembles must calibrate its boundary-aware upper limit, false-alarm probability, look-elsewhere effect, and signal-dependent self-noise. For injected amplitude $A_{\rm inj}$, the calibration diagnostics are
\begin{equation}
 \begin{aligned}
 b_A&=\frac{\langle\widehat A_{\rm GW}\rangle-A_{\rm inj}}{\sigma_A},
 &v_A&=\frac{{\rm Var}(\widehat A_{\rm GW})}{\sigma_A^2},\\
 c_{95}&={\rm Pr}(A_{\rm inj}\le A_{95}).
 \end{aligned}
 \label{eq:injection_diagnostics}
\end{equation}
Unbiased weak-signal recovery has $b_A=0$ and $v_A=1$, while a calibrated 95\% construction has $c_{95}=0.95$.  These quantities are evaluated across injected amplitudes, spectral shapes, station/target subsets, and deliberate noise-model mismatch before a discovery power is quoted.

The conventional covariance $\bm C_0$ must include heteroscedastic photon noise from the measured HCCR throughput, colored atmosphere and oscillator noise, hardware-delay and AM-to-PM calibration, station coordinates and loading, Earth orientation, reflector coordinates and thermal phase-center motion, libration, tides, and dynamical-model error.  Deterministic HCCR throughput sets the heteroscedastic photon covariance and accepted-window selection.  Calibrated HCCR phase or range templates enter the deterministic model and its design matrix; uncertain template amplitudes and residual red processes enter $\bm C_0$.  A throughput harmonic becomes a range-line nuisance only through an unremoved amplitude-to-phase transfer.  An omitted coherent or correlated term can bias $\widehat A_{\rm GW}$ and is not made safe merely by increasing $\Dcov$.

The decisive checks are therefore fixed in advance: a candidate covariance must follow the common Earth--Moon response across stations and resolved targets; exhibit the predicted range--range-rate relation; be suppressed in pure reflector differences by the factor $B/\LEM$; remain stable under station, target, elevation, wavelength, and epoch jackknives; and be recovered with correct coverage in blind injections into deliberately mismatched noise and dynamical models.  With the same sampled kernel and standardized critical threshold, the weak-signal Fisher approximation to Eq.~(\ref{eq:dcovdef}) is $\Dcov\simeq(F_{AA}^{\rm white,fixed}/F_{AA}^{\rm full,marg})^{1/2} =(q_{\rm noise}q_{\rm det}q_{\rm hyp})^{-1/2}$; calibrated boundary or injection limits need not obey this equality exactly.  Realized timestamps, measured HCCR transfer functions, physical ephemeris partials, and colored-noise covariances determine the numerical $\Dcamp$ and $\Dcov$.  The present unit-factor curves therefore serve as direct commissioning requirements for an end-to-end calibrated SGWB likelihood.

\emph{Conclusion.}--High-power AM-CW LLR constitutes a quantitative SGWB experiment based on coherent absolute ranging to individually resolved single HCCRs.  It measures the phase of a GHz modulation envelope and propagates the resulting range observable directly into a time-domain likelihood for the GW-driven mean and covariance of the lunar orbit.  Near $\fii=0.847245\,\uHz$, a broadband SGWB produces accumulated orbital drift and a covariance growing as $t^3$, giving the experiment a sharply defined orbital-harmonic response rather than an instantaneous baseline-strain observable.

The scalar 1-kW reference link gives $5.893\times10^3\,\mathrm{s}^{-1}$ and an $88.8$--$124.3\,\mu$m photon contribution in 100 s.  The projected $50\,\mu$m campaign requires a $30\,\mu$m photon term, combined non-shot noise no larger than $40\,\mu$m, and $1500$ statistically useful windows yr$^{-1}$.  Even eliminating that photon term would improve the covariance-branch threshold by at most $(50/40)^2=1.56$.  HCCR phase-center calibration, atmospheric and hardware-delay stability, orbital-response sampling, and parameter covariance therefore control the realized SGWB sensitivity once the photon requirement is met.

For a broadband covariance-dominated response, $\Ogw^{95}\propto\sigma_R^2\nuEff^{-1}T_{\rm obs}^{-4}$.  The response-anchored five-year 95\% thresholds are $1.86\times10^{-7}\Dtot$ for $80\,\mu$m and $500\,\mathrm{yr}^{-1}$ and $2.42\times10^{-8}\Dtot$ for $50\,\mu$m and $1500\,\mathrm{yr}^{-1}$.  In the latter case, $\Dtot\le1.15$ gives nominal source-statistic equality for the $\Ogw=7.08\times10^{-8}$ compact-binary benchmark; the phase-transition benchmark requires $\Dtot\le0.92$.  Transfer-function measurements, realized-timestamp likelihoods, information-contraction tests, and blind injections make these conditions falsifiable commissioning criteria.  Because it measures accumulated lunar orbital dynamics rather than pulsar pulse-arrival times or direct interferometric strain, AM-CW LLR occupies a distinct response space and supplies an independent GW measurement channel.  High-power AM-CW LLR therefore has a defined five-year exclusion sensitivity and a testable route to SGWB detection at $0.85\,\uHz$ on the basis of its own measured performance.

\emph{Acknowledgments.}--The work described here was carried out at the Jet Propulsion Laboratory, California Institute of Technology, under contract with the National Aeronautics and Space Administration.

\bibliographystyle{apsrev4-2}

%

\end{document}